\documentstyle[aps,epsfig,prl,floats]{revtex}

\newcommand{\be}{\begin{equation}}
\newcommand{\ee}{\end{equation}}
\newcommand{\bea}{\begin{eqnarray}}
\newcommand{\eea}{\end{eqnarray}}

\newcommand{\naco}{Na$_x$CoO$_2\cdot y$H$_2$O}

\tightenlines
\begin{document}
\draft

\title{Order and disorder in the triangular-lattice $t-J-V$ model
at 2/3 electron density}

\author{
        Weihong Zheng$^{(a)}$,
	      Jaan Oitmaa$^{(a)}$,
        Chris J.~Hamer$^{(a)}$, and
        Rajiv R.~P.~Singh$^{(b)}$
}
\address{
$^{(a)}$ School of Physics, University of New South Wales, Sydney NSW
2052, Australia\\
$^{(b)}$ Department of Physics, University of California, Davis, CA
95616\\
}

%\author{Weihong Zheng}
%\email[]{w.zheng@unsw.edu.au}
%\homepage[]{http://www.phys.unsw.edu.au/~zwh}
%\affiliation{School of Physics,
%The University of New South Wales,
%Sydney, NSW 2052, Australia.}
%
%\author{C.J. Hamer}
%\email[]{c.hamer@unsw.edu.au}
%\affiliation{School of Physics,
%The University of New South Wales,
%Sydney, NSW 2052, Australia.}
%
%\author{R.R.P. Singh}
%\email[]{singh@raman.ucdavis.edu}
%\affiliation{Department of Physics, University of California, Davis,
%CA95616, USA}

\twocolumn[\hsize\textwidth\columnwidth\hsize\csname
@twocolumnfalse\endcsname

\date{\today}
\maketitle

\begin{abstract}
Motivated by the recent discovery of superconductivity in
\naco,
we use series expansion methods and cluster mean-field theory
to study spontaneous charge
order, N\'eel order, ferromagnetic order, dimer order and phase-separation
in the triangular-lattice $t-J-V$ model at $2/3$ electron density.
We find that for $t<0$, the charge ordered state, with electrons
preferentially occupying a honeycomb lattice, is very robust.
Quite surprisingly, hopping to the third sublattice can even enhance
N\'eel order.
% to increase.
At large
negative $t$ and small $V$, the Nagaoka ferromagnetic state is obtained.
%, but other singlet ground states remain close in energy.
For large positive $t$, charge and N\'eel order
vanish below a critical $V$, giving rise to
an itinerant antiferromagnetically correlated state.
\end{abstract}

\pacs{PACS numbers: 71.27.+a, 73.43.Nq, 71.10.Fd}
%\keywords{}

%71.27.+a Strongly correlated electron systems; heavy fermions
%73.43.Nq Quantum phase transitions
%71.10.Fd Lattice fermion models (Hubbard model, etc.)
%75.40.Gb Dynamic properties (dynamic susceptibility, spin waves, spin diffusion, dynamic scaling, etc.)
%75.10.Jm Quantized spin models
%75.50.Ee Antiferromagnetics

]

\narrowtext

The recent discovery of superconductivity in the
Na${}_x$CoO${}_2 \cdot y$H${}_2$O
 materials has led to considerable theoretical
excitement \cite{baskaran,shastry,lee}.
Even though the superconducting
transition temperatures in these materials are not as high as in the cuprate
based high temperature superconductors, a primary reason for
the theoretical excitement is that the underlying lattice-geometry for the
spin-1/2 cobalt sites in these materials is triangular. The antiferromagnetic
exchange on the triangular-lattice is frustrated and has long been argued
to provide the right conditions for exotic spin-physics in terms of resonating
valence bonds (RVB) \cite{anderson,laughlin}. While most recent studies of the
triangular-lattice Heisenberg model ground states \cite{lhuillier,huse}
suggest a magnetically ordered phase, RVB physics could still be very important
away from half-filling.

The materials \naco  have a very complex phase-diagram with doping $x$
and water concentration $y$ that includes superconducting, insulating,
charge ordered, and also magnetic and phase separated behavior \cite{cava}.
Superconductivity is found only in a
narrow range of $x$-values (in the approximate range $1/4<x<1/3$)
and the transition temperatures ($T_c$) have
dome-like structures with $x$, with a maximum in the middle and vanishing
at the ends \cite{cava2}. Sodium is primarily believed to be just
an electron donor to the cobalt oxide layers, where much of the electronic
activity takes place. Thus an interpretation of the $T_c$
variation is that at certain commensurate electron fillings $T_c$ goes to
zero and the system becomes a Mott insulator, and superconductivity
arises from doping the Mott insulator \cite{baskaran,lee}.
%The role of water has been
%argued in these materials to provide screening for
%the charges thus reducing the effective intermolecular coulomb repulsion
%felt by the electrons and stabilizing the doping values.

%With variations in both $x$ and $y$, these materials provide
%an interesting interplay of superconducting, insulating, charge ordered,
%and also magnetic and phase-separated behavior.
%The stoichiometric material CoO$_2$ has cobalt ions with valence $+4$.
%This is a $d^5$ configuration carrying spin-half. On the other end the material
%NaCoO$_2$ has cobalt ions
%with valence $+3$, which would be a non-magnetic $d^6$ configuration.
%Because superconductivity occurs near $x=1/3$,
%theoretical interest has been particularly high in a state where $2/3$
%of the cobalt atoms have spin-half.
%

In this paper, we wish to focus on the origin and nature of insulating
behavior at $2/3$ electron density, which may be appropriate
for Na$_x$CoO$_2$ at $x=1/3$, near the superconducting region.
There have been very few numerical studies of $t-J$ or large-$U$ Hubbard
models on the triangular-lattice.
Quantum Monte Carlo methods are known
to suffer from minus sign problems, while the exact diagonalization methods
are limited to very small systems.
Here we use series expansion methods \cite{gel00}, which are especially appropriate for
addressing the Mott insulating behavior. We study the Hamiltonian:
\bea
H &=& -t \sum_{\langle ij\rangle} P ( c_{i\sigma}^\dag c_{j\sigma} + {\rm h.c.} ) P
   + J \sum_{\langle ij\rangle} ( {\bf S}_i \cdot {\bf S}_j - {n_i n_j\over 4} )
   \nonumber \\
  &&+ V \sum_{\langle ij\rangle}  (1-n_i) (1-n_j)
\eea
The first and second terms are the usual terms for the
$t-J$ model, while the last term with coupling $V$ is a
nearest-neighbor  hole-hole repulsion term.
The on-site repulsion is assumed to be infinite and no double occupancy is allowed.

For large $V$,
the electrons will spontaneously charge order, preferentially
occupying two of the three sublattices of the triangular lattice.
This is equivalent to a fully occupied honeycomb lattice.
The honeycomb-lattice Heisenberg model has
been studied before \cite{weihong} and, not surprisingly,
the unfrustrated model shows antiferromagnetic N\'eel order. An interesting
question is whether the
ability of the electrons to hop to the third sublattice can
destabilize the antiferromagnetic order or promote ground states
with different symmetries. Our results show that for small $V$ a number
of different ground state phases, including charge ordered, N\'eel ordered,
dimerized, ferromagnetic, phase separated and short-range
antiferromagnetic phases can arise.

%By developing Ising type
%expansions around charge-ordered and N\'eel-ordered ground states, we study the
%extent of N\'eel order in these systems. By developing columnar dimer expansions,
%we study tendencies for dimerization and resonating valence bond ground states.
%Both types of expansions allow us to study the extent of spontaneous charge
%order. Comparing the energies of the Heisenberg model on triangular and
%honeycomb lattices, we find that electronic phase separation, between
%fully occupied triangular-lattice and empty regions occurs over a very
%narrow range of parameters for positive small $V$ and small $|t|$.
%Ferromagnetism is studied
%by comparing to the energies of the fully polarized Nagaoka state.

We find that the sign of the hopping matrix element $t$ plays a significant
role in the properties of the system. First consider
negative $t$. In this case, charge order
is very robust, and for a substantial range of $|t|/J$ values,
extends even to $V=0$. Also
%For large $V$,
increasing $|t|$ causes the magnetization on the honeycomb
sites to increase well beyond the value in the pure Heisenberg model.
For $V=J=0$, the Nagaoka ferromagnetic state is the ground state.
However, singlet, possibly dimerized phases compete with the Nagaoka
state even at very small $V$ and $J$ values.
%For larger $J/|t|$ and $V=0$ a N\'eel ordered
%state competes for the ground state with other singlet states.
These results support the earlier
high temperature expansion results
of Koretsune and Ogata \cite{ogata} who found evidence for substantial low
temperature entropy in the $t-J$ models with negative $t$.

For positive $t$, hopping reduces N\'eel and charge order.
Below a critical $V$ spontaneous charge and N\'eel order vanish.
From the series expansion one cannot
conclusively show that the two vanish simultaneously,
but
%they happen so close to each other that it
the numerics
is consistent with
a single transition.
At this transition, the symmetry between the three sublattices
is restored both with respect to occupancy and hopping.
Also, the dimerized and singlet states are never favored for positive $t$.
Thus, for small $V$ and large $t$, there appears to be a phase transition to
an itinerant state with short-range antiferromagnetic correlations.

%We have done both Ising expansion and
%Columnar dimer expansion to study this system.
%The linked cluster series
%expansion method has been previously reviewed in
%several articles,\cite{gel00} and will not be repeated here.

\begin{figure}
  \begin{center}
    \begin{minipage}[t]{0.5\linewidth}
      \epsfig{file=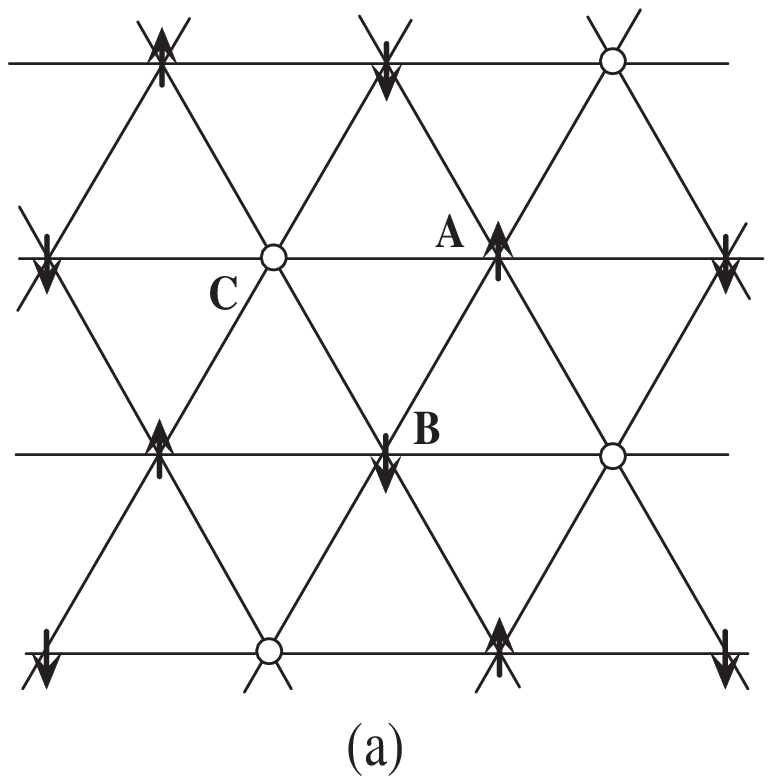, width=\linewidth}
    \end{minipage}\hfill
    \begin{minipage}[t]{0.5\linewidth}
      \epsfig{file=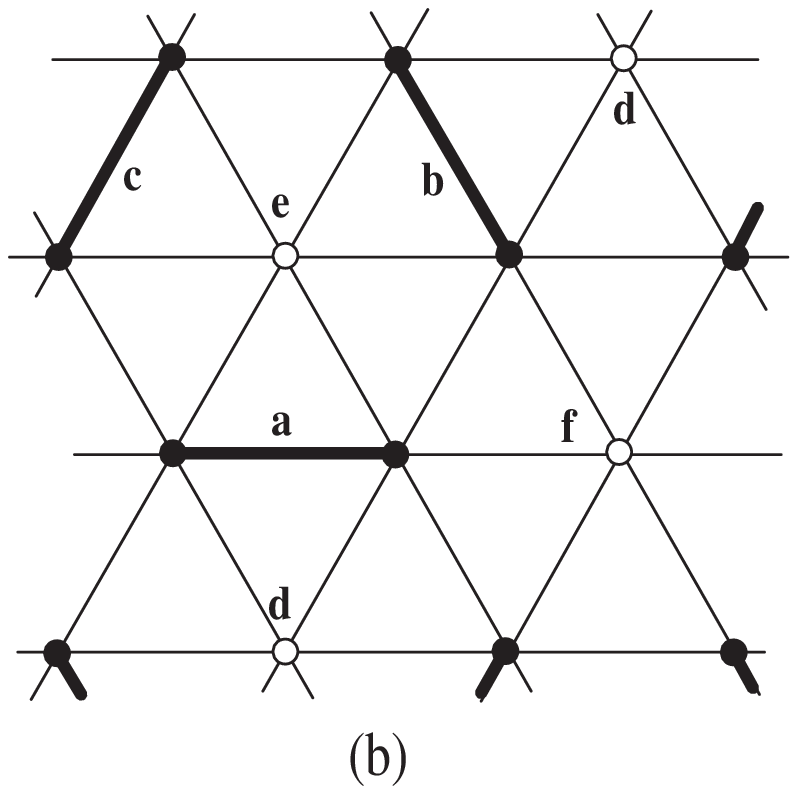, width=\linewidth}
    \end{minipage}
  \end{center}
      \caption{(a) The N\'eel ordered state in the Ising limit,
      (b) the columnar dimer ordered state. The symbols denote the various distinguishable
      sublattices. \label{fig1}}
\end{figure}

To perform the Ising expansion, we divided the Hamiltonian into an unperturbed Hamiltonian ($H_0$)
and a perturbation ($H_1$) as follows,
\bea
H &=& H_0 + \lambda H_1  \\
H_0 &=& \sum_{\langle ij\rangle} [ J ( S_i^z S_j^z - \frac{n_i n_j}{4} )
+ V (1-n_i) (1-n_j) ] \nonumber \\
&& + U {\Big [}  - \sum_{i\in A} S^z_i + \sum_{i\in B} S^z_i + \sum_{i\in C} n_i {\Big ]}
   \\
H_1 &=& \sum_{\langle ij\rangle} [ J ( S_i^x S_j^x + S_i^y S_j^y )
    - t P ( c_{i\sigma}^\dag c_{j\sigma} + {\rm h.c.} )P ] \nonumber \\
&& - U {\Big [}  - \sum_{i\in A} S^z_i + \sum_{i\in B} S^z_i + \sum_{i\in C} n_i {\Big ]}
\eea
where $\lambda$ is an expansion parameter.
Note that we have divided the lattice into three sublattices ($A$, $B$ and $C$,
as shown in Fig. 1(a),
The last term in both
$H_0$ and $H_1$ is a local field term on three sublattices, which can be included
to improve convergence. The limits $\lambda=0$ and $\lambda=1$
correspond to the Ising model and the original
model, respectively. The  unperturbed ground state is
the usual Ne\'el state (shown in Fig. 1(a)).
The Ising series have been calculated for various ground state properties
% the ground-state
% energy per site $E_0/N$, the staggered magnetization $M$ on $A$ or $B$ sublattices,
% (note that the $\langle S_i^z \rangle$ on $C$ sublattice is always zero: this means,
% in principle,
% that charge order is not required to have spin order), the electron density $n$ on $A$,
% $B$ and $C$ sublattices (note we always have
%  $n_A=n_B$), the nearest
% neighbor spin-spin correlation function $\langle {\bf S}_i \cdot {\bf S}_j \rangle $, and
% the lowest lying triplet excitation energies $\Delta (k_x ,k_y)$
for several ratios of the couplings and (simultaneously) for several
values of $U$ to order $\lambda^{11}$.
The calculation involves a list of 231955 linked clusters (up to 10 sites)
for a triangular lattice with 3 sublattices.
% and 6 symmetries (3 3-fold rotation symmetries and 3 reflection symmetries)
The  series are available on request. Note that $\langle S_i^z \rangle$ on the $C$
sublattice is always zero: this means,
in principle, that spin order can occur without charge order.

\begin{figure}
  \begin{center}
      \epsfig{file=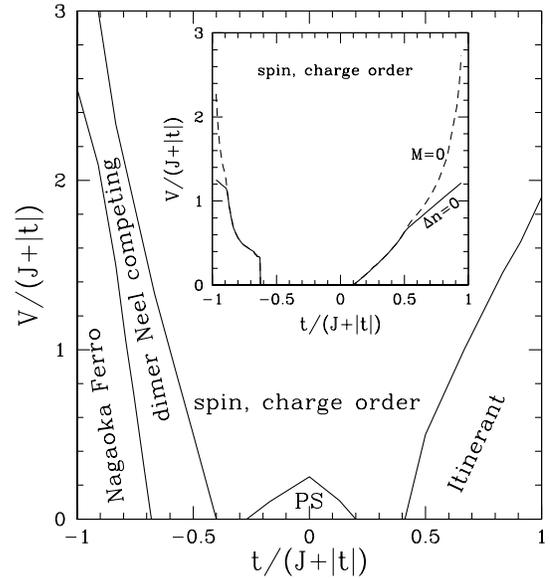, width=0.9\linewidth}
  \end{center}
      \caption{
      The phase diagram for the model. PS referes to the fully
		phase-separated region. Inset shows the phase diagram
      given by 3-sites cluster mean field calculation.
%(remark: it seem to have a N\'eel order down to $V=0$ even for $t>0$, see Fig. 4.)
\label{fig2}}
\end{figure}

We also perform a dimer series expansion starting from the columnar dimer pattern as shown in
Fig. 1(b), where we take the couping for bold bonds (the ``dimer pairs") to be $J$ and
the coupling for thin bonds (the ``free pairs") to be $\lambda J$. The
electron hopping amplitude (on all bonds) is also taken to be $\lambda t$, i.e. we divided the Hamiltonian
into an unperturbed Hamiltonian ($H_0$)
and perturbation ($H_1$) as follows:
\bea
H &=& H_0 + \lambda H_1  \\
H_0 &=& J \sum_{\rm dimer~pairs} ({\bf S}_i \cdot {\bf S}_j - \frac{n_i n_j}{4} )
+ V \sum_{\langle ij\rangle} (1-n_i) (1-n_j) \nonumber \\
&& + U \sum_{\rm empty~sites} n_i
% + {\rm (Env. of V-term)}
 \\
H_1 &=& J \sum_{\rm free~pairs} ( {\bf S}_i \cdot {\bf S}_j - \frac{n_i n_j}{4} )
- t \sum_{\langle ij\rangle} P ( c_{i\sigma}^\dag c_{j\sigma} + {\rm h.c.} ) P \nonumber \\
&& - U \sum_{\rm empty~sites} n_i
\eea
where again $\lambda$ is an expansion parameter.
The last term in both
$H_0$ and $H_1$ is a local field term on the empty sites (i.e.,
the sublattices $d$, $e$ and $f$ shown
in Fig. 1(b)), which can be included
to improve convergence.
% Again the operator $H_0$ is taken as the unperturbed
% Hamiltonian, with
The unperturbed ground state is
the product state of dimer singlets.
To perform this expansion, we have to divide the lattice into
6 sublattices: 3 for 3 different oriented dimers, and 3 for 3 different
empty sites (i.e. the sublattices $d$, $e$ and $f$ shown
in Fig. 1(b)), and this makes it quite difficult to generate
the cluster data for this expansion. The dimer series has been computed
 to order $\lambda^{8}$ for various ground state properties and for
the excitation spectrum, for several
ratios of the couplings and (simultaneously) for several
values of $U$ \cite{later}.

\begin{figure}
  \begin{center}
      \epsfig{file=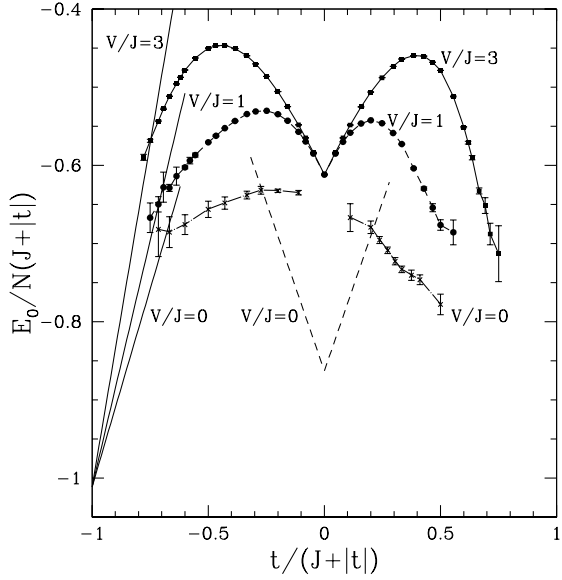, width=0.9\linewidth}
  \end{center}
      \caption{The ground state energy
      vs $t/(J+|t|)$ for $V/J=0,1,3$.
      The points with error bars are the results of the Ising expansion, while
      the straight lines near $t/(J+|t|)=-1$
      are the ground state energies, i.e., $E_0/N =1.01138 t + 0.2481 V$,
       for the ferromagnetic state \protect\cite{ferro-energy}.
The straight lines near $t/(J+|t|)=0$
      are the energies  for a fully phase separated state, i.e
      $ E_0/N = 2 J e^{\rm tri}/3+V-J/2 = -0.862 J + V $, where
      $e^{\rm tri}=-0.543$ is the ground state energy per site for the
      Heisenberg antiferromagnet on the triangular lattice\protect\cite{huse}. \label{fig3}}
\end{figure}

Some of the results are qualitatively confirmed by a cluster mean-field
theory. Here a triangle of 3-spins ($A$, $B$ and $C$ as shown in Fig.1(a))
is considered in the effective field of its surroundings,
which favors antiferromagnetic N\'eel order on two of the sublattices.
In other words, we consider the following Hamiltonian
\bea
 H &=& J ( {\bf S}_{\mbox{{\tiny $A$}}} \cdot {\bf S}_{\mbox{{\tiny $B$}}}
 + {\bf S}_{\mbox{{\tiny $B$}}} \cdot {\bf S}_{\mbox{{\tiny $C$}}}
 + {\bf S}_{\mbox{{\tiny $C$}}} \cdot {\bf S}_{\mbox{{\tiny $A$}}}  )
 - t ( c_{{\mbox{{\tiny $A$}}}\sigma}^\dag c_{{\mbox{{\tiny $B$}}}\sigma} + \nonumber \\
 &&  c_{{\mbox{{\tiny $B$}}}\sigma}^\dag c_{{\mbox{{\tiny $C$}}}\sigma}
 + c_{{\mbox{{\tiny $C$}}}\sigma}^\dag c_{{\mbox{{\tiny $A$}}}\sigma}
 + {\rm h.c.} )  - h_s (S_{\mbox{{\tiny $A$}}}^z - S_{\mbox{{\tiny $B$}}}^z ) + \mu_{\mbox{{\tiny $V$}}}
\eea
where the fields $h_s = 2 J M$, $\mu_{\mbox{{\tiny $V$}}} =  V [
  \tilde{n}_{\mbox{{\tiny $C$}}} ( 3 n_{\mbox{{\tiny $A$}}} + 3 n_{\mbox{{\tiny $B$}}} )
+ \tilde{n}_{\mbox{{\tiny $A$}}} ( 3 n_{\mbox{{\tiny $B$}}} + 3 n_{\mbox{{\tiny $C$}}} )
+ \tilde{n}_{\mbox{{\tiny $B$}}} ( 3 n_{\mbox{{\tiny $A$}}} + 3 n_{\mbox{{\tiny $C$}}} )
] $. $H$ needs to be diagonalized with the following  self-consistent
conditions
\bea
M &=& \langle S_{\mbox{{\tiny $A$}}}^z \rangle = - \langle S_{\mbox{{\tiny $B$}}}^z \rangle,
\quad  \langle S_{\mbox{{\tiny $C$}}}^z \rangle =0  \\
n_{\mbox{{\tiny $A$}}} &=& \langle \tilde{n}_{\mbox{{\tiny $A$}}} \rangle
= n_{\mbox{{\tiny $B$}}} = \langle \tilde{n}_{\mbox{{\tiny $B$}}} \rangle, \quad
n_{\mbox{{\tiny $C$}}} = \langle \tilde{n}_{\mbox{{\tiny $C$}}} \rangle
\eea

\begin{figure}
  \begin{center}
      \epsfig{file=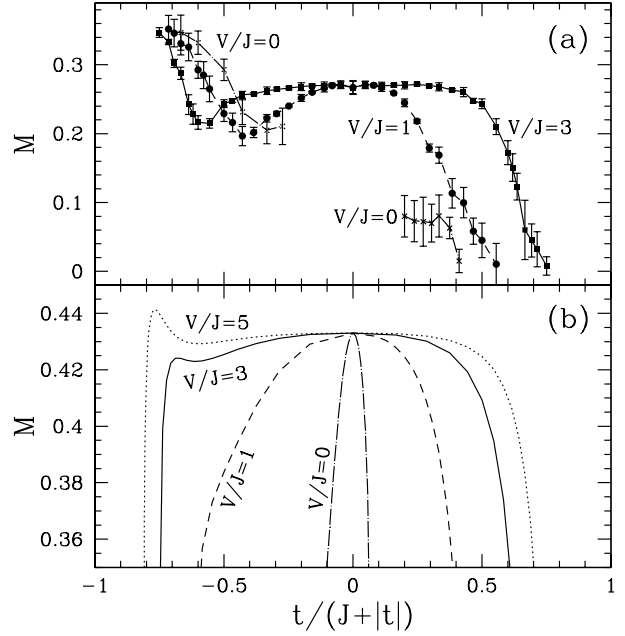, width=\linewidth}
  \end{center}
      \caption{The magnetization $M$ on sublattice $A$ vs $t/(J+|t|)$ for $V/J=0,1,3$, obtained from
      Ising expansions (a) and a 3-site cluster mean-field calculation (b).\label{fig4}}
\end{figure}

\begin{figure}
  \begin{center}
      \epsfig{file=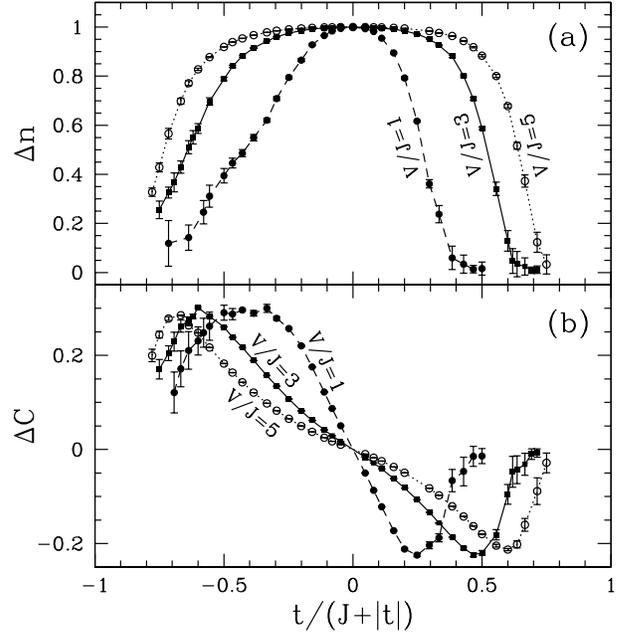, width=\linewidth}
  \end{center}
      \caption{The difference of electron density on sublattice $A$ (or $B$) and $C$
      $\Delta n = n_{\mbox{{\tiny $A$}}}-n_{\mbox{{\tiny $C$}}}$
       and hopping amplitudes
       $\Delta C = \langle c_{\mbox{{\tiny $A$}}\sigma}^\dag c_{\mbox{{\tiny $C$}}\sigma} + h.c. \rangle_0 -
      \langle c_{\mbox{{\tiny $A$}}\sigma}^\dag c_{\mbox{{\tiny $B$}}\sigma} + h.c. \rangle_0 $    vs $t/(J+|t|)$ for $V/J=1,3,5$,
      obtained from Ising expansions.\label{fig5}}
\end{figure}

The complete phase diagram for the model is shown in Fig.~2. The inset shows the
phase diagram obtained by cluster mean-field theory. To obtain the
regions of ferromagnetism, we need to compare
our calculated energies, with those of the Nagaoka state
\cite{nagaoka,ferro-energy,muller}. These are shown in Fig.~3.
The phase separated (PS) region is obtained by
comparing the calculated energies with those of a phase separated state,
with a complete separation of holes and spins. These comparisons are also
shown in Fig.~3.
Comparison of ground state energies from dimer and N\'eel expansions
leads us to conclude a competing dimer/N\'eel region adjacent to the
Nagaoka phase. This may also reflect the onset of incommensurate
spin-correlations, which are not explored here.

The key results of the
mean-field theory are that:
(i) for negative $t$, hopping can increase
N\'eel order on the occupied sublattices, and (ii) for positive $t$ and large $J$ charge
and spin order vanish simultaneously.
For $J=0$, $t>0$, this mean-field calculation does not fully
capture the N\'eel order at large $V$ resulting from $t/V$ perturbations.
Thus, for small $J$, charge
and spin order vanish at separate critical $V$'s.

In Fig.~4, we show the variation of the N\'eel order parameter with
hopping. Fig.~4(a) shows the series expansion results and Fig.~4(b) the
results of cluster mean-field theory. The qualitative resemblance
is clear. For positive $t$ and small $V$ there is a rapid monotonic
decrease in N\'eel order with $t$. For negative $t$, increasing
$|t|$ leads to a non-monotonic behavior and an increase in the
N\'eel order parameter beyond that of the Heisenberg model. This
non-intutive result is confirmed by the cluster mean-field theory
and reflects an interference phenomenon. Fig.~5 shows the
restoration of charge order symmetry between the sublattices in the
Ising expansions. Fig. 5(a) shows the difference between the site occupations
between the initially occupied and unoccupied sublattices. Fig. 5(b)
shows the asymmetry in effective nearest-neighbor
hopping amplitudes between the different
sublattices. The latter shows a non-monotonic behavior with $t$. However,
our results are consistent with the restoration of both types of
sublattice symmetry at the same positive $t$, whereas the symmetry
is not restored for negative $t$. These results suggest that below
a critical $V$ and for positive $t$ the system goes into an
itinerant antiferromagnetic state with short-range magnetic
correlations through a continuous phase transition, whereas
for $t<0$, there is a first order transition to a Nagaoka state.

In conclusion, we have shown in this paper that the triangular-lattice
$t-J-V$ model with $2/3$ electron density has a very rich phase diagram
that consists of charge ordered, N\'eel, ferromagnetic, dimerized,
phase separated and short-range antiferromagnetically correlated regions.
Clearly the full phase-diagram of the triangular-lattice $t-J$ model
with doping will be extremely rich, with the possibility of Mott-insulating
phases at many commensurate dopings. While it is difficult to directly
relate these calculations to the \naco materials, our results
support the idea that superconductivity in these materials may be viewed
as arising from doping a Mott insulator at $x=1/3$ provided that
$t>0$ \cite{dsingh,pickett}. One could further
speculate that the water content can change the effective $V$ and bring the system
closer to or across the insulating-itinerant phase boundary, thus playing
a major role in the onset of superconductivity.

%\section{Acknowledgement}
We would like to thank W. E. Pickett and R. T. Scalettar for
many useful discussions.
This work is supported by a grant
from the Australian Research Council and by US National Science Foundation
grant number DMR-0240918.
 We are grateful for the computing resources provided
 by the Australian Partnership for Advanced Computing (APAC)
National Facility and by the
Australian Centre for Advanced Computing and Communications (AC3).

\bibliography{basename of .bib file}

\begin{references}
%\bibitem[*]{zwh} Email address: w.zheng@unsw.edu.au
%\bibitem[\dag]{cjh} Email address: c.hamer@unsw.edu.au

\bibitem{baskaran} G. Baskaran, Phys. Rev. Lett. {\bf 91}, 097003 (2003);
cond-mat/0310241,cond-mat/0306569.

\bibitem{shastry} B. Kumar and B. S. Shastry, Phys. Rev. B{\bf 68},
104508 (2003).

\bibitem{lee} Q.-H. Wang, D.-H. Lee and P. A. Lee, cond-mat/0304377;
O. I. Motrunich and P. A. Lee, cond-mat/0310387 and cond-mat/0401160.

\bibitem{anderson} P. W. Anderson, Mater. Res. Bull. {\bf 8}, 153 (1973);
P. Fazekas and P. W. Anderson, Philos. Mag. {\bf 30}, 423 (1974).

\bibitem{laughlin} V. Kalmeyer and R. B. Laughlin, Phys. Rev. Lett.
{\bf 59}, 2095 (1987).

\bibitem{lhuillier} B. Bernu {\it et. al.} Phys. Rev. B{\bf 50},
10048 (1994).

\bibitem{huse} R. R. P. Singh and D. A. Huse, Phys. Rev. Lett. {\bf 68},
1766 (1992).

\bibitem{cava} Y. J. Uemura {\it et. al.} cond-mat/0403031;
H. W. Zandbergen {\it et. al.} cond-mat/0403206;
Q. Huang {\it et. al.} cond-mat/0402255; M. L. Foo {\it et al}
cond-mat/0312174.

\bibitem{cava2} R. E. Schaak {\it et. al.} Nature {\bf 424} 527 (2003).

\bibitem{gel00}M.P. Gelfand and R.R.P. Singh, Adv. Phys. {\bf 49}, 93(2000).

\bibitem{weihong} J. Oitmaa, C.J. Hamer and W. Zheng, Phys. Rev. B{\bf 45},
9834 (1992).

\bibitem{ogata} T. Koretsune and M. Ogata, Phys. Rev. Lett. {\bf 89},
116401 (2002).

\bibitem{later} Details will be presented elsewhere.

\bibitem{nagaoka} Y. Nagaoka, Phys. Rev. {\bf 147}, 392 (1966).

\bibitem{ferro-energy} B. S. Shastry, H. R. Krishnamurthy and P. W. Anderson,
Phys. Rev. B{\bf 41}, 2375 (1990).

\bibitem{muller} T. Hanish, B. Kline, A. Ritzi and E. Muller-Hartmann,
Ann. Phys. (N.Y.) {\bf 4}, 303 (1995).


\bibitem{dsingh} D.J. Singh, Phys. Rev. B{\bf 61}, 13397(2000).

\bibitem{pickett} For an alternative viewpoint see
K.-W. Lee, J. Kunes and W. E. Pickett, cond-mat/0403018.

\end{references}

% \newpage

%=======================================================================
\end{document}